\documentclass[a4paper]{article}

\usepackage{INTERSPEECH2020}
\usepackage{hyperref}
\usepackage{spconf,amsmath,graphicx,url, amssymb}
\usepackage{cite}

\usepackage{comment}
\usepackage{hyperref}

\title{Phase aware speech enhancement using Realisation of Complex-valued LSTM}
\name{Raktim Gautam Goswami, Sivaganesh Andhavarapu and K Sri Rama Murty}
\address{Speech Information Processing Lab, Department of Electrical Engineering\\
Indian Institute of Technology Hyderabad, Hyderabad - 502285, INDIA\\
\texttt{[ee17btech11051, ee18resch11020, ksrm]@iith.ac.in}}

\begin{document}

\maketitle
\begin{abstract}
Most of the deep learning based speech enhancement (SE) methods rely on estimating the magnitude spectrum of the clean speech signal from the observed noisy speech signal, either by magnitude spectral masking or regression. 
These methods reuse the noisy phase while synthesizing the time-domain waveform from the estimated magnitude spectrum.  
However, there have been recent works highlighting the importance of phase in SE.
There was an attempt to estimate the complex ratio mask taking phase into account using complex-valued feed-forward neural network (FFNN). 
But FFNNs cannot capture the sequential information essential for phase estimation. 
In this work, we propose a realisation of complex-valued long short-term memory (RCLSTM) network to estimate the complex ratio mask (CRM) using sequential information along time.
The proposed RCLSTM is designed to process the complex-valued sequences using complex arithmetic, and hence it preserves the dependencies between the real and imaginary parts of CRM and thereby the phase. 
The proposed method is evaluated on the noisy speech mixtures formed from Voice-Bank corpus and DEMAND database.\footnote{Audio samples are available at (best viewed in google chrome) \url{https://siplabiith.github.io/rclstm.html}}
When compared to real value based masking methods, the proposed RCLSTM improves over them in several objective measures including perceptual evaluation of speech quality (PESQ), in which it improves by over 4.3\%.
\end{abstract}


\section{Introduction}
\label{sec:intro}

Speech enhancement (SE) refers to the process of improving the intelligibility and quality of degraded speech by the separation of additive noise from it. In many day to day tasks such as telecommunication and speech recognition, a major step in preprocessing the input signal is removing unwanted noise from it. We can also find SE being used in hearing aids, voice analysis and speaker identification. Despite its many uses, removing noise to extract the clean speech is still considered as one of the most challenging tasks in speech processing.

Historically, many approaches have been taken for enhancing the quality of speech. Some of the early works include spectral subtraction \cite{C38}, Wiener filtering \cite{C2}, Statistical model-based methods \cite{C31}, subspace algorithms \cite{C22}, Time-Frequency (TF) Mask estimation using STFT \cite{C32, C28}. Although these methods give good results with high SNR  and stationary noise conditions, these are not very effective in low SNR conditions and non-stationary noises.

The success of deep learning and neural networks has motivated researchers to apply it to the task of SE. DNN based method can't make any explicit assumption on noise type and noise conditions, these methods try to estimate the speech characteristics. DNN based models \cite{C24} and auto encoder architectures \cite{C7,C30} have successfully shown that neural networks work much better than traditional methods when the SNR of noisy file is low. This led researchers to extend this idea to the use of recurrent neural networks (RNNs) \cite{C8, C10}, especially LSTM. While most of these methods worked with speech in the frequency domain, some approaches~\cite{C40, C20} were able to produce good results in the time domain as well.

%The results by Valentini et al.\cite{C10} show that RNNs work better than DNNs. This can be attributed to the fact that RNNs are more suited in handling sequential information like speech signals compared to DNNs.

%In most of the neural network based approaches, the spectral magnitude of noisy speech is taken as input and that of the clean speech is taken as the labels. STFT of a signal is always in complex domain and thus contain a magnitude and a phase. However, in
Most of the neural network based approaches, use the magnitude of the STFT of the noisy speech to predict the magnitude of the STFT of the clean speech. The noisy phase is reused to reconstruct the target speech signals. This is being done as it was claimed that phase is unimportant for SE~\cite{C4}. But, recent studies ~\cite{C5} have shown that speech quality and intelligibility can be further increased when a clean phase is used for reconstruction.
While many approaches have used the Cartesian distance between the predicted signal and clean signal as the objective function, there has also been significant work in incorporating loss functions that directly improve the objective measures (\cite{pesqloss, monaural}).

Williamson et al.\cite{C28} have demonstrated that using a complex-valued ratio mask (CRM) performs better than real-valued ratio mask in objective results such as perceptual evaluation of speech quality (PESQ), short  time  objective  intelligibility (STOI) and segmental signal to noise ratio
(SSNR).
In the proposed model, we consider both the real and imaginary parts of the STFT of noisy speech to predict the real and imaginary parts of the CRM. We perform a complex-valued multiplication of the input with this CRM to get the enhanced output. For the selection of hidden layer size, we followed the principles of subspace-based SE method~\cite{C22}. The underlying assumption made was that in a $k$-dimensional noisy speech signal, the clean speech signal does not span entire $k$-dimensional space but instead spans a smaller dimension $m$ $(m\leq{k})$.
The advantages of using our method can be summarised as 
\begin{itemize}
    \item The network operates in the complex domain, thus using both magnitude and phase information simultaneously.
    \item The complex-valued network is realised using real LSTM cells with complex multiplication between the outputs of each cell, which is suitable for processing complex valued data like STFT coefficients.
    \item The network is used to predict the CRM for the complex valued STFT of the noisy speech signal. 
\end{itemize}
The rest of the paper is organized as follows: Definition of complex ratios mask, unbounded nature of mask and counter measures are discussed in Section~\ref{sub:compmask}. The building blocks of RCLSTM, network architecture and objective functions are discussed in section~\ref{sec:netarch}. In Section~\ref{sec:exp}, experimental evaluation of the proposed method and comparison with the existing state-of-the-art SE algorithms are discussed. Section~\ref{sec:conc} summarizes the important contributions of this work and highlights possible future directions for improvement.

\section{Complex ratio mask (CRM)}
\label{sub:compmask}
Let $x[n]$ be the noisy speech signal collected in a noisy environment,
\begin{equation}
    x[n]=s[n]+v[n],
\end{equation}
where $s[n]$ is the clean speech signal and $v[n]$ is the additive background noise.
As speech is a non-stationary signal, the effect of additive noise is not uniform across time and frequency. 
As a result, the noisy speech signal exhibits varying SNR across time frames and varying sub-band SNR across frequency bins~\cite{C29}.
In order to deal with the varying degrees of degradation in the TF plane, noisy signal is typically processed in the STFT domain~\cite{C23}.
Let $S[n,k]$ and $X[n,k]$ denote the STFT of the clean speech signal $s[n]$ and noisy speech signal $x[n]$, respectively, where $n=1,2,\ldots N$ are the time-frames and $k=1,2,\ldots K$ are the frequency bins. 
DNN based SE methods aim at learning mask in the TF plane to retrieve an estimate of the clean signal $\hat s[n]$ from $x[n]$ in the STFT domain, i.e.,
\begin{equation}
   \hat S[n,k]= M[n,k] X[n,k]
    \label{eq:speech_estimate}
\end{equation}
where $M[n,k]$ is referred to as the mask to be applied on the noisy speech signal. 

The mask is intended to preserve the high SNR regions in the TF plane and suppress the low SNR regions.
Several types of masks have been proposed in the literature for SE~\cite{C24}.
In the case of ideal binary mask (IBM)~\cite{C25}, $M[n,k]$ is assigned a 1 or 0 depending on the signal power $|S[n,k]|^2$ is higher than the noise power $|V[n,k]|^2$ or not in that particular TF bin.
Hence ideal binary mask can be interpreted as creating a hard threshold of $X[n,k]$.
Ideal ratio mask (IRM), on the other hand, provides a soft threshold by multiplying $X[n,k]$ with a continuous value in the range $[0,1]$ given by $\sqrt{\frac{|S[n,k]|^2}{|S[n,k]|^2+|V[n,k]|^2}}$. 
Both IBM and IRM require explicit knowledge of the additive noise $v[n]$ for mask computation.
Hence, these masks cannot be applied in cases where $v[n]$ is not known, e.g. cellular transmission.  
Although SMM~\cite{C26,sig}, defined as $\frac{|S[n,k]|}{|X[n,k]|}$, does not require $v[n]$ it may result in unbounded values.
It should be noticed that these masks scale only the magnitude component of $X[n,k]$ while retaining the noisy phase as it is.

There have been several studies highlighting the importance of phase in speech perception, especially in noisy environments~\cite{C27}. 
Complex masks have been proposed in the literature to account for the the effect of phase~\cite{C28}.
In this paper we use the CRM proposed in~\cite{C28} for SE.
In this approach, the CRM is defined as 
\begin{equation}
    \label{eq:crm}
    M[n,k]=\frac{S[n,k]}{S[n,k]+V[n,k]} = \frac{S[n,k]}{X[n,k]}.
\end{equation}
The real and imaginary parts of the mask can be expressed as 
\begin{eqnarray}
    M_r[n,k] & = &\frac{X_r[n,k] S_r[n,k] + X_i[n,k] S_i[n,k]}{X_r^2[n,k] + X_i^2[n,k]} \\ 
    M_i[n,k] & = &\frac{X_r[n,k] S_i[n,k] - X_i[n,k] S_r[n,k]}{X_r^2[n,k] + X_i^2[n,k]}
    \label{eq:mask_real_imag}
\end{eqnarray}
where $Z_r$ and $Z_i$ denote the real and  and imaginary parts of a complex number $Z$. 
In this work, we proposed RCLSTM network to estimate the real and imaginary parts of the CRM $M[n,k]$ from the noisy signal $x[n]$. The estimated mask can be used to retrieve the clean speech signal using~(\ref{eq:speech_estimate}).

\subsection{Bounding the dynamic range of CRM}
\label{ssub:opt_mask}
In general, the CRM in (\ref{eq:crm}) can range from $(-\infty,+\infty)$. 
Although most of the values of the mask $M[n,k]$ are in acceptable range, it is observed that a few of the TF bins exhibit very large values in the mask.
The mask $M[n,k]$ in~(\ref{eq:crm}) takes very large values for those bins where the STFT of the noisy speech signal $X[n,k]$ is close to zero. 
In such TF bins, the magnitude and phase of the clean signal and noise are related by
\begin{eqnarray*}
     |S[n,k]| &\approx&  |V[n,k]| \\ 
     \angle{S[n,k]} &\approx& \pi+\angle{V[n,k]}.
\end{eqnarray*}
That is, in certain bins the magnitudes of the speech signal and additive noise are approximately equal and their phases differ approximately by $\pi$ radians.
Even though the SNR is 0 dB in such bins, the signal cannot be retrieved as the observed noisy signal is zero in those bins because of the phase difference.
If we try to retrieve these TF bins, it will lead to unnecessary artifacts in the enhanced signal.
Hence, instead of retrieving those TF bins by retaining large values of the mask, we propose to compress the dynamic range of the mask using hyperbolic tangent $(\tanh)$ function. 
In this work, the hyperbolic tangent function is applied individually on the real and imaginary parts of the CRM to arrive at a bounded CRM given by
\begin{equation}
    B[n,k]=\tanh(M_R[n,k])+j\tanh(M_I[n,k])
\end{equation}
Unlike CRM $M[n,k]$, the bounded CRM $B[n,k]$ has its real and imaginary parts restricted to the range $[-1,1]$, and hence it is better suited for modeling. 
%We propose to use a deep complex-valued LSTM network to estimate the bounded CRM $B[n,k]$ from the STFT of the noisy signal $X[n,k]$.

\section{CRM estimation using Realisation of Complex valued LSTM}
\label{sec:netarch}
For enhancing a noisy speech signal $x[n]$, we need to estimate the bounded CRM $B[n,k]$ from the STFT of the noisy speech signal $X[n,k]$. 
In~\cite{C24}, a FFNN was used to estimate the CRM from the STFT of the noisy speech signal. 
Since FFNN is a memoryless system, it is not adequate to capture the long-term temporal dependencies across the speech frames. 
LSTM, on the other hand, is capable of modeling long-temporal dependencies across the speech frames.  
Unlike FFNNs, LSTM has recurrent connections through which encapsulated past memory is propagated to learn the sequential information~\cite{C33}. 
LSTMs have been successfully used to model real-valued sequences in several application including, but not limited to, speech processing~\cite{C34}, handwriting recognition \cite{C35}, stock market prediction etc.
In this work, we extend the LSTM to process complex-valued sequences. 

\begin{figure}[t]
\begin{minipage}[b]{1.0\linewidth}
  \centering
  \centerline{\includegraphics[width=0.7\linewidth]{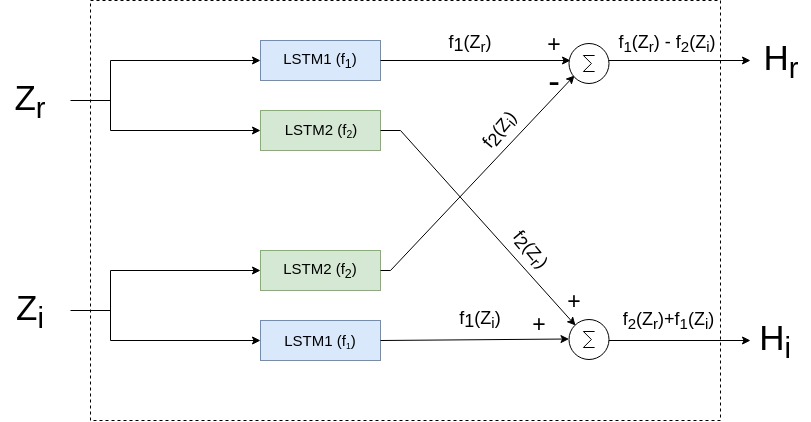}}
\end{minipage}
\caption{Composing a RCLSTM from a pair of real-valued LSTMs}
\vspace{-0.6cm}
\label{fig:comp_lstm}
\end{figure}

Let $\mathbf{Z}[t]\in \mathbb{C}^P$, $1\le t \le T$ denote a complex-valued sequence of $P$-dimensional vectors over $T$ time-steps. We need to design a realisation of complex-valued LSTM for sequence-to-sequence mapping of complex-vectors. 
Let $H[t] \in \mathbb{C}^Q$, $1 \le t \le T$ denote the sequence of $Q$-dimensional output vectors over the $T$ time-steps. 
In this work, we used a pair of real-valued LSTMs to compose a realisation of complex valued LSTM as shown in~\ref{fig:comp_lstm}. 
In the proposed architecture for the RCLSTM, the real and imaginary parts of the output sequence $\mathbf{H}[t]$ are computed in a similar manner as complex multiplication, and are given by
\begin{eqnarray}
      \mathbf{H}_r[t] = f_1(\mathbf{Z}_r[1:t]) - f_2(\mathbf{Z}_i[1:t]) \quad 1 \le t \le T \\
      \mathbf{H}_i[t] = f_2(\mathbf{Z}_r[1:t]) + f_1(\mathbf{Z}_i[1:t]) \quad 1 \le t \le T 
\end{eqnarray}
where $f_1:\mathbb{R}^P \rightarrow \mathbb{R}^Q$ and $f_2:\mathbb{R}^P \rightarrow \mathbb{R}^Q$ denote the nonlinear mappings of LSTM-1 and LSTM-2, respectively.   

\subsection{Network architecture}
In order to estimate the bounded CRM $B[n,k]$ from $X[n,k]$, we used a stack of two RCLSTM layers and a dense layer~\cite{C12} as shown in Fig.~\ref{fig:model_block}.
In this architecture, the bounded mask $B[n,k]$ in each TF-bin is estimated using a context of 21 frames around $X[n,k]$.
The actual time-steps and dimensions of each layer are given in Fig.~\ref{fig:model_block}.
Notice that the second layer of RCLSTM returns output only at the last time-instant.
The complex-valued outputs of the second layer are then passed thorough the complex-valued dense layer with $\tanh(.)$ activation functions. 
%The network weights are adjusted to minimize the error between the desired and estimated masks.
\subsection{Objective function}
\label{ssub:Obj}
In this work, $L_2$ norm of the error between the desired and the estimated masks is used as an objective measure to quantify the goodness of the estimated mask.   
The averaged norms of the errors across a random batch is used as an objective function to optimize the network weights. 
The objective function for a particular random batch $\mathcal{S}$ is given by
\begin{equation}
    \label{eq:obj}
    J = \frac{1}{|\mathcal{S}|}\sum_{j \in \mathcal{S}}{\sum_{k=1}^{k=P}{\left(\hat B[j,k] - B[j,k]\right) \left(\hat B[j,k] - B[j,k]\right)^*}}
\end{equation}
where $|\mathcal{S}|$ denotes the cardinality of the random batch $\mathcal{S}$, $(.)^*$ denotes the complex conjugate, and $P$ is the number of frequency bins in the mask. The weights of the network are updated using stochastic gradient descent algorithm to minimize the objective function.

\section{Experimental Evaluation}
\label{sec:exp}
The performance of the proposed RCLSTM in enhancing noisy speech signals is evaluated on a data set composed by Valentini et al.~\cite{C10}, which is freely available online\footnote{{\url{https://datashare.is.ed.ac.uk/handle/10283/2791}}}.
This data set is prepared by combining clean speech samples from 30 speakers with real-time noise recordings from DEMAND database~\cite{C15} at different SNRs.
The train set consists of speech data from 28 speakers (14 male + 14 female) mixed with 10 types of noises at 4 SNRs, 0, 5, 10 and 15 dB. 
The test set consists of speech data from 2 speakers (1 male + 1 female) mixed with 5 types of noises at 4 SNRs: 2.5, 7.5, 12.5 and 17.5 dB.  
The speakers and noises types in train and test sets are completely non-overlapping.
There are around 20 different sentences from each speaker under every noise condition.
All the speech signals are downsampled to 16 kHz for further processing.
\begin{figure}[t]
\begin{minipage}[b]{1.0\linewidth}
  \centering
  \centerline{\includegraphics[width=8.5cm]{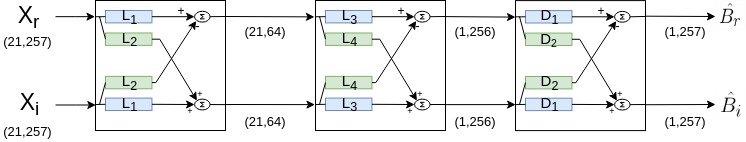}}
\end{minipage}
\caption{Block diagram of proposed model, which takes context dependent frames as input and estimates real and imaginary parts of the mask. }
\vspace{-0.5cm}
\label{fig:model_block}
\end{figure}

The STFT of signal is computed with frame size of 512 samples (32 ms) shifted by 256 samples (16 ms). Due to complex conjugate nature of the STFT coefficients, only the 257 non-negative frequency bins are retained for further processing. 
The real and imaginary parts of the CRM for each TF bin are calculated using~(\ref{eq:crm}) and are bounded to range $[-1, +1]$ using $\tanh()$ compression. 
A network of RCLSTMs, shown in Fig.~\ref{fig:model_block} is used to estimate the bounded CRM $B[n,k]$ from the STFT of noisy speech signal $X[n,k]$. 
The network consists of two RCLSTM layers and a dense layer with 64, 257 and 257 nodes, respectively.
In order to take the temporal context into account while estimating the bounded CRM, the input is presented over a context of 21 frames (10 left and 10 right). 
The exact dimensions of the outputs at each layer are indicated in Fig.~\ref{fig:model_block}. 
The network is trained to minimize the objective function in~(\ref{eq:obj}) using {\it adam} optimizer~\cite{C36}.
Once the network is trained, it is used to estimate the bounded CRM $B[n,k]$ from the STFT of noisy signal $X[n,k]$. 
The CRM $M[n,k]$ is computed from $B[n,k]$ using inverse hyperbolic tangent function and is multiplied with $X[n,k]$ as in~(\ref{eq:speech_estimate}) to obtain an estimate of the STFT of clean speech signal $\hat S[n,k]$.
An estimate clean speech signal $\hat s[n]$ is reconstructed from $\hat S[n,k]$ using overlap and add method~\cite{C37}.
\subsection{Objective evaluation}
The performance of the proposed RCLSTM based SE method is evaluated using the following objective measures. A higher value of these measures mean better enhancement. All of these have been calculated using open-source implementation\footnote{\url{https://www.crcpress.com/downloads/K14513/K14513_CD_Files.zip}}.
\begin{itemize}
    \item PESQ: Perceptual evaluation of speech quality \cite{C17}. Range in $(-0.5,4.5)$
    \item SSNR: Segmental SNR \cite{C18}. Range in $(0,\infty) $
    \item CSIG, CBAK, CMOS: Mean opinion score (MOS) of predictor of signal distortion, background-noise intrusiveness and overall signal quality respectively~\cite{C18}. Range in $(1,5)$
    %\item CBAK: MOS predictor of background-noise intrusiveness~\cite{C18}. Range in $(1,5)$
    %\item CMOS: MOS predictor of overall signal quality \cite{C18}. Range in $(1,5)$
    %\item STOI: Short time objective intelligibility \cite{C19}. Range in $(0,1)$
\end{itemize}

These objective measures were used to fix the number of nodes in the first RCLSTM layer. Table~\ref{tab:nodes_comp} compares performance of the proposed method with 32, 64, and 128 nodes in the first RCLSTM layer. 
The performance is optimal when the input is projected on 64-dimensional subspace , and hence we used 64 nodes for rest of the experiments in the paper. 
It should be noticed that a very small subspace leads to loss of speech-specific information, while a larger subspace forward propagates the noise as well. 
\begin{table}[h]
    \centering
    \vspace{-0.5cm}
    \caption{Effect of hidden-layer dimension on performance } 
\begin{tabular}{ |c|c|c|c| } 
         \hline
        \textbf{Nodes in first LSTM layer} & \textbf{32}  & \textbf{64} & \textbf{128} \\
        \hline
        \textbf{PESQ} & 2.60 & \textbf{2.62} & 2.58 \\
        \hline
        %\textbf{SSNR} & 7.87 & 8.01 & \textbf{8.35} \\
        %\hline
        %\textbf{CSIG} & 3.76 & \textbf{3.80} & 3.76 \\
        %\hline
        %\textbf{CBAK} & 3.17 & 3.18 & \textbf{3.19} \\
        %\hline
        %\textbf{CMOS} & 3.17 & \textbf{3.18} & 3.16 \\
        % \hline
        \end{tabular}
    \label{tab:nodes_comp}
\end{table}

\begin{table}[h]
    \centering
        \vspace{-0.5cm}
        \caption{Relative improvement PESQ across SNRs \& genders}
        \begin{tabular}{ |c|c|c|c|c|c|c| } 
         \hline
        \textbf{SNR(dB)} & \textbf{17.5} & \textbf{12.5} & \textbf{7.5} & \textbf{2.5} & \textbf{Male} & \textbf{Female}\\
         \hline
         \hline
        \textbf{Noisy} &2.6  &2.1  &1.76  &1.42  & 2.17 & 1.80\\
         \hline
        \textbf{Enhanced} & 3.16 & 2.79 & 2.50 & 2.03 & 2.67 & 2.62\\
         \hline
         \textbf{Impr.(\%)} & 21 & 32.8 & 42 & 42.9 & 23 & 45.5\\
         \hline
        \end{tabular}
    \label{tab:snr_compn}
    \vspace{-0.5cm}
\end{table}
%\vspace{-0.5cm}
\begin{table}[h]
    \centering
    \caption{Masking based methods compassion with RCLSTM} 
\begin{tabular}{ |c|c|c|c|c| } 
         \hline
        \textbf{Model} & \textbf{SMM~\cite{sig}}& \textbf{CIRM~\cite{sig}}  & \textbf{RILSTM} & \textbf{RCLSTM} \\
        \hline
        \textbf{PESQ} &2.51 &2.49 & 2.55 & \textbf{2.62} \\
        \hline
       
        %\textbf{CSIG} & \textbf{3.89} & 3.73 & 3.69 & 3.80  \\
        %\hline
        %\textbf{CBAK} & 2.44 & 2.51 &  2.48& \textbf{3.18}\\
        %\hline
        %\textbf{CMOS} & 3.20 & 3.10 & 3.10&  \textbf{3.23}\\
         %\hline
        %  \textbf{STOI} &0.93  & 0.92& 0.91& 0.93  \\
        % \hline
        \end{tabular}
    \label{tab:obj_comp_mask}
    \vspace{-0.3cm}
\end{table}

\begin{figure}[h!]
\centering

  \includegraphics[ width=0.8\columnwidth]{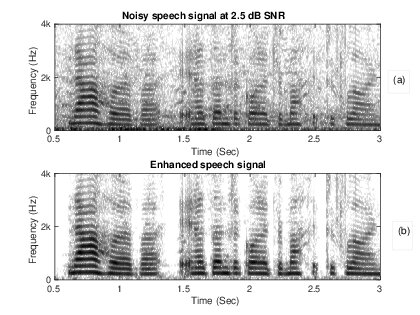}
\vspace{-0.3cm}
\caption{Spectrograms of noisy and enhanced speech signals from a female speaker - p257-411 in the test set}
\vspace{-0.2cm}
\label{fig:spectrogram1}
\end{figure}

\begin{table}[h]
    \centering
    \caption{Comparison of RCLSTM with the state-of-the-art methods}
        \begin{tabular}{ |c|c|c|c|c|c| } 
        \hline
        \textbf{Model} & Noisy & Weiner & SEGAN & M-GAN & \textbf{RCLSTM}  \\
        \hline
        \hline
        \textbf{PESQ} & 1.97 &2.22 & 2.16 & 2.53  & \textbf{2.62} \\
        \hline
        \textbf{SSNR} &  1.68 & 5.07 & 7.73 & -   & \textbf{8.01}\\
        \hline
        \textbf{CSIG} & 3.35 & 3.23 & 3.48 & \textbf{3.80}  & \textbf{3.80}\\
        \hline
        \textbf{CBAK} & 2.44 & 2.68 & 2.94 & 3.12  & \textbf{3.18}\\
        \hline
        \textbf{CMOS} & 2.63 & 2.67 & 2.80 & 3.14  & \textbf{3.23} \\
        \hline
       % \textbf{STOI} & 0.91 & -  & \textbf{0.93} & \textbf{0.93} & \textbf{0.93}\\
       % \hline
        \end{tabular}
    \label{tab:obj_comp}
    \vspace{-0.4cm}
\end{table}
%\vspace{-1cm}
Table~\ref{tab:snr_compn} shows the effect of SNR and gender on the proposed enhancement method using RCLSTMs. The PESQ metric consistently improved at all SNRs illustrating the effectiveness of the proposed SE method. It can also be observed that the proposed method performed equally well on both male and female speakers. 
The lower PESQ metric of noisy speech signals from female speakers can be attributed to their higher fundamental frequency. 
In the case of female speakers, pitch harmonics occur with wider gaps between them in the narrow-band spectrogram~\cite{C39}, as shown in Fig.~\ref{fig:spectrogram1}.
Because of these wider gaps between the pitch harmonics, the effect of noise is higher on the perceptual quality of female speakers than the male speakers.
The enhancement algorithms should be able to clean the noise embedded in between the pitch harmonics to  achieve good performance on female speakers. 
Fig.~\ref{fig:spectrogram1} illustrates the effectiveness of the proposed method in cleaning the noise in between the pitch harmonics, and thereby achieving a relative  improvement of over 45\% in PESQ metric for the speech sample.

%\vspace{-2cm}
\begin{figure}[h!]
\centering
  \vspace{-0.3cm}
  \includegraphics[width=0.8\columnwidth]{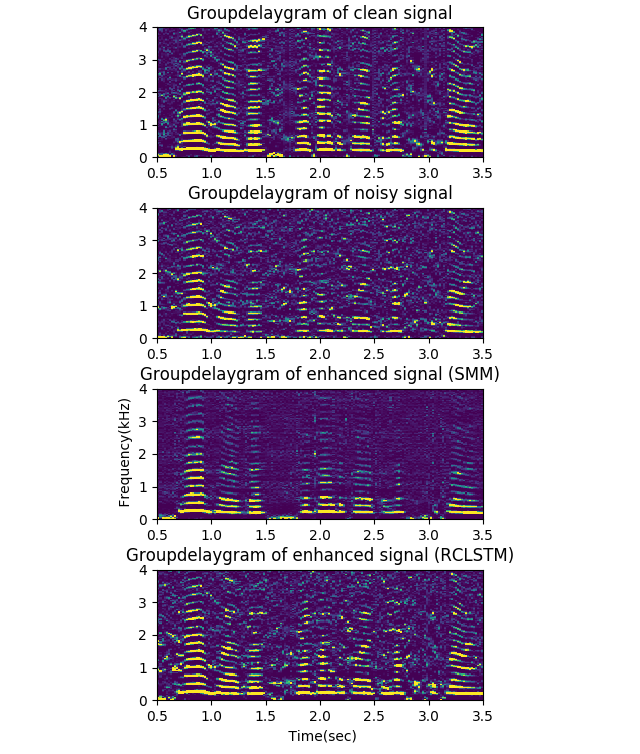}
\vspace{-0.5cm}
\caption{ Groupdelaygrams of clean, noisy and enhanced speech signals from a female speaker - p257-411 in the test set}
\vspace{-0.6cm}
\label{fig:gd1}
\end{figure}
%

%
%\begin{table*}[b]
 %   \centering
 %   \caption{Objective measures of enhanced speech signal comparison with the state-of-the-art method}
  %      \begin{tabular}{ |c|c|c|c|c|c|c|c| } 
 %       \hline
  %      \textbf{Model} & Noisy & Weiner\cite{C2} & SEGAN\cite{C20} & MMSE-GAN\cite{C21} & Magnitude-based LSTM & Real-Img & \textbf{RCLSTM}  \\
   %     \hline
   %     \hline
   %     \textbf{PESQ} & 1.97 &2.22 & 2.16 & 2.53 & 1.19 & 2.55 & \textbf{2.62} \\
     %   \hline
     %   \textbf{SSNR} &  1.68 & 5.07 & 7.73 & - & -0.86 & -2.53 & \textbf{8.01}\\
    %    \hline
    %    \textbf{CSIG} & 3.35 & 3.23 & 3.48 & \textbf{3.80} & 2.49 & 3.69 & \textbf{3.80}\\
    %    \hline
   %     \textbf{CBAK} & 2.44 & 2.68 & 2.94 & 3.12 & 1.71 & 2.48 & %\textbf{3.18}\\
      %  \hline
       % \textbf{CMOS} & 2.63 & 2.67 & 2.80 & 3.14 & 1.74 & 3.10 & \textbf{3.19} \\
      %  \hline
      %  \textbf{STOI} & 0.91 & -  & \textbf{0.93} & \textbf{0.93} & 0.73 & 0.91 & \textbf{0.93}\\
      %  \hline
       % \end{tabular}
    %\label{tab:obj_comp}
%\end{table*}
%
Table~\ref{tab:obj_comp_mask} provides a comparison of proposed RCLSTM-based CRM method with the other masking methods, viz., SMM~\cite{sig}, complex ideal ratio mask (CIRM)~\cite{sig} and another model consisting of two separate networks for each of the real and imaginary parts of the complex masks (RILSTM) on the same data-set. 
The SMM model uses the noisy spectral magnitude to estimate the clean spectral magnitude and re-uses the noisy phase for the reconstruction of the signal back into the time domain. 
CIRM model uses noisy spectral magnitude to estimate real and imaginary parts of the complex mask.
% The limitation of SMM and CIRM is that the input features of the network are only the log spectral magnitudes. Because of this, performance is inferior to methods that use both input and output as complex values.
The main limitation of SMM and CIRM is that these models ignore the phase information of the signal during enhancement.
% In case of CIRM, the noisy spectral magnitude is used to estimate the real and imaginary parts of complex mask. 
%Since, it uses real valued noisy spectral magnitude as input and predicts the outputs its performance is inferior to methods that use both input and output as complex values.
In another experiment (RILSTM), we used two separate networks for estimating the real and imaginary parts of the mask separately and later combined them during prediction of the enhanced signal. However, there is no correlation between the real and imaginary parts of the signal in this model. RCLSTM resolves the limitations of the methods mentioned above and therefore improves over them in objective measures.
The groupdelaygrams in Fig(\ref{fig:gd1}) illustrate the effectiveness of the proposed RCLSTM in enhancing the pitch harmonics. 
As compared to magnitude based enhancement method (SMM\cite{sig} in Fig(\ref{fig:gd1})), RCLSTM is seen to preserve the pitch harmonics much better, especially at frequencies above 2 $kHz$.
% The pitch harmonics are much better preserved in the frequencies above 2 $kHz$ in the groupdealygram of  RCLSTM compared with the magnitude based enhancement method (SMM).   
This further supports our claim that the RCLSTM network enhances both the magnitude and phase of the noisy speech signal.
 
Table~\ref{tab:obj_comp} provides a comparison of the proposed RCLSTM-based enhancement method with existing state-of-the-art methods, viz., SEGAN~\cite{C20} and MMSE-GAN~\cite{C21}. 
We have included the adaptive signal processing based Wiener filtering~\cite{C2} approach also for comparison.
All these three methods aim at enhancing the magnitude spectrum of the speech signal. The phase spectrum from noisy speech signal is reused while reconstructing the time-domain signal. 
On the other hand, the proposed RCLSTM estimates the CRM using both the magnitude and phase of the noisy speech signal.
It can be observed that the performance of the proposed method is either better than or as good as the existing approaches across all the objective measures. 

\section{Conclusion}
\label{sec:conc}
%\vspace{-0.2cm}
In this work, we developed RCLSTM to jointly model the magnitude and phase of the complex-valued temporal sequences. 
The realisation of complex-valued LSTM is composed using two-real valued LSTMs which are interconnected through complex-arithmetic. 
The RCLSTM is used to estimate the complex ratio mask from the STFT coefficients of noisy speech signal. 
The estimated CRM is multiplied with the STFT of noisy speech signal to retrieve the clean speech signal. 
The proposed approach performed better than the existing magnitude-based enhancement algorithms in both objective and subjective measures.
These results illustrate ability of RCLSTM in effectively modeling the phase.
Possible future works can include changing the gate structure of LSTMs to incorporate complex numbers, and extending the application of RCLSTM to audio source separation and dereverberation.

\bibliographystyle{IEEEtran}

\bibliography{refs}

% \begin{thebibliography}{9}
% \bibitem[1]{Davis80-COP}
%   S.\ B.\ Davis and P.\ Mermelstein,
%   ``Comparison of parametric representation for monosyllabic word recognition in continuously spoken sentences,''
%   \textit{IEEE Transactions on Acoustics, Speech and Signal Processing}, vol.~28, no.~4, pp.~357--366, 1980.
% \bibitem[2]{Rabiner89-ATO}
%   L.\ R.\ Rabiner,
%   ``A tutorial on hidden Markov models and selected applications in speech recognition,''
%   \textit{Proceedings of the IEEE}, vol.~77, no.~2, pp.~257-286, 1989.
% \bibitem[3]{Hastie09-TEO}
%   T.\ Hastie, R.\ Tibshirani, and J.\ Friedman,
%   \textit{The Elements of Statistical Learning -- Data Mining, Inference, and Prediction}.
%   New York: Springer, 2009.
% \bibitem[4]{YourName17-XXX}
%   F.\ Lastname1, F.\ Lastname2, and F.\ Lastname3,
%   ``Title of your INTERSPEECH 2020 publication,''
%   in \textit{Interspeech 2020 -- 20\textsuperscript{th} Annual Conference of the International Speech Communication Association, September 15-19, Graz, Austria, Proceedings, Proceedings}, 2020, pp.~100--104.
% \end{thebibliography}

\end{document}